\documentclass[fleqn,usenatbib]{mnras}

\usepackage{newtxtext,newtxmath}

\usepackage[T1]{fontenc}

\DeclareRobustCommand{\VAN}[3]{#2}
\let\VANthebibliography\thebibliography
\def\thebibliography{\DeclareRobustCommand{\VAN}[3]{##3}\VANthebibliography}

\def\gcc{\hbox{\rm\hskip.35em  g cm}$^{-3}$}

\def\radss{\hbox{\rm\hskip.35em  rad s}$^{-2}$}

\usepackage{graphicx}	
\usepackage{amsmath}	

\title[Pulse profile with glitch of PSR J1048$-$5832]
{Pulse profile variability associated with the glitch of PSR J1048$-$5832}

\author[P. Liu et al.]{
P. Liu,$^{1,2}$
J.-P. Yuan,$^{2,3,4}$\thanks{E-mail: yuanjp@xao.ac.cn}
M.-Y. Ge,$^{5}$\thanks{E-mail: gemy@ihep.ac.cn}
W.-T. Ye,$^{3,5}$
S.-Q. Zhou,$^{6}$
S.-J. Dang,$^{7}$
Z.-R. Zhou,$^{8,9}$
 \newauthor
E. G{\"u}gercino{\u{g}}lu,$^{10}$\thanks{E-mail: egugercinoglu@gmail.com}
W.-H. Wang,$^{11}$
P. Wang,$^{10,12}$
A. Li,$^{1}$\thanks{E-mail: liang@xmu.edu.cn}
D. Li,$^{13,10,14}$
and N. Wang$^{2,4,15}$
\\
$^{1}$Department of Astronomy, Xiamen University, Xiamen, Fujian 361005, China\\
$^{2}$Xinjiang Astronomical Observatory, Chinese Academy of Sciences, 150 Science 1-Street, Urumqi, Xinjiang 830011, China\\
$^{3}$University of Chinese Academy of Sciences, Chinese Academy of Sciences, Beijing 100049, China\\
$^{4}$Xinjiang Key Laboratory of Radio Astrophysics, 150 Science1-Street, Urumqi, Xinjiang, 830011, China\\
$^{5}$Key Laboratory of Particle Astrophysics, Institute of High Energy Physics, Chinese Academy of Sciences, Beijing 100049, China\\
$^{6}$School of Physics and Astronomy, Sun Yat-Sen University, Zhuhai, 519082, China\\
$^{7}$School of Physics and Electronic Science, Guizhou Normal University, Guiyang 550001, China\\
$^{8}$National Time Service Center, Chinese Academy of Sciences, Xi'an, 710600, China\\
$^{9}$Key Laboratory of Time Reference and Applications, Chinese Academy of Sciences, Xi'an, 710600, China\\
$^{10}$National Astronomical Observatories, Chinese Academy of Sciences, 20A Datun Road, Chaoyang District, Beĳing 100101, China\\
$^{11}$College of Mathematics and Physics, Wenzhou University, Wenzhou 325035, China\\
$^{12}$Institute for Frontiers in Astronomy and Astrophysics, Beijing Normal University, Beijing 102206, China\\
$^{13}$Department of Astronomy, Tsinghua University, Beijing 100084, China\\
$^{14}$Computational Astronomy Group, Zhejiang Laboratory, Hangzhou 311100, China\\
$^{15}$Key Laboratory of Radio Astronomy, Chinese Academy of Sciences, Urumqi, Xinjiang, 830011, China
}
\date{Accepted XXX. Received YYY; in original form ZZZ}
\pubyear{2024}

\begin{document}
\label{firstpage}
\pagerange{\pageref{firstpage}--\pageref{lastpage}}
\maketitle

\begin{abstract}
PSR J1048$-$5832 (B1046$-$58) is a Vela-like pulsar that has exhibited multiple glitch events. 
In this study, we analyze the timing data spanning nearly 16 years, acquired from both the Fermi Gamma-ray Space Telescope and the Parkes 64 m radio telescope.
As a result, a total of five glitches are detected within this dataset.
Among them, a previously unknown small glitch is newly found at MJD 56985(9) (November 24, 2014), making it the smallest glitch recorded from this source so far. 
The increments of the spin frequency and its first derivative are $\Delta \nu \approx 2.2(3) \times 10^ {-8} $ Hz, and $\Delta \dot{\nu} \approx 3(2) \times 10^ {-15}$  s$^{-2}$, respectively.
Significant changes in the integrated normalized mean pulse profile are detected following three of the five glitch events, notably in the radio band. 
Although no evidence of a correlation is found between the spin-down rate and profile evolution, the jump phenomenon of $W_{55}$ (pulse width at the 55\% peak amplitude) after the glitch in the narrow mode suggests that the glitch may influence the profile change.
We discuss the influence of glitches on the pulsar's emission properties in terms of platelet motion by a crustquake and also put constraints on the equation of state from the moment of inertia and response timescales of involved superfluid layers inside the neutron star.
\end{abstract}



\section{Introduction}

Pulsars are highly magnetized neutron stars that exhibit rapid and stable rotation. 
By employing the pulsar timing technique \citep{EdwardsHM2006}, the difference between the time of arrival (ToAs) of the pulse signal at the telescope and the ToAs estimated by the model can be analyzed to obtain the timing residuals.
Long-term timing observations reveal that pulsar glitch (sudden and temporary changes) \citep{RadhakrishnanM1969, Reichley1D969} is one of the irregular phenomena that primarily influence the spin evolution of pulsars.
Thus, glitch observations have significant implications for our understanding of pulsar physics~\citep{HaskellM2015,2016ApJS..223...16L,2021ApJ...923..108S}. 
Most interestingly, glitches can impact the pulsar's emission properties, including its pulse profile and spectral characteristics. Studying how glitches affect emission can provide insights into the radiation processes near the pulsar's surface, as well as the internal dynamics~\citep{2019AIPC.2127b0004Y,AntonopoulouHE2022}.
Apart from glitches, irregularities in the rotation of pulsars are usually described as timing noise. Like glitches, timing noise is mostly observed in young pulsars; Unlike glitches, timing noise manifests as "red" noise processes, where timing residuals are dominated by low-frequency fluctuations. The primary source of intrinsic red noise over long timescales is believed to arise from a random walk in the pulse phase, spin frequency (or angular velocity), and spin-down rate (or torque)~\citep{BoyntonGHN1972}.

\begin{table*}
\caption{The parameters and data span of PSR J1048$-$5832. The parameters include the right ascension and declination, the spin period and its first derivative
e, the surface dipole magnetic field ($B_{\rm s} = 3.2 \times 10^{19} \sqrt{P\dot{P}} \,{\rm G}$), the characteristic age ($\tau_{\rm c} = P/(2\dot{P})$), and the dispersion measure.}
\label{Tab:1}
\vspace{-0.3cm}
\begin{center}
\begin{tabular}{ccccccccccc}
  \hline
      \hline    
Pulsar Name   & RA       & DEC & $P$  & $\dot{P}$  & $B_{\rm s}$  & Age  & DM &  Fermi-LAT Span  & Parkes Span\\          
(PSR)         &(hh:mm:ss)& ($ ^{\degr} : ^{\arcmin} : ^{\arcsec}$) & (s) & (10$^{-15}$)   & ($10^{12}$ G) & (kyr) & (cm$^{-3}$pc) & (MJD)  & (MJD)\\  
\hline      
J1048$-$5832$^{1}$  &10:48:13.1(1)$^{2}$& $-$58:32:03(1)$^{2}$   &0.123725$^{3}$  &96.1217$^{3}$  &3.49  &20.4   & 128.679$^{4}$   &54704 -- 59977   &54302 -- 59182 \\  
\hline 
\end{tabular}
\item[]  References for parameters of the pulsar: 1 \citep{JohnstonSMK1992}; 2  \citep{LowerBSJ2020}; 3 \citep{JankowskiBSK2019}; 4 \citep{PetroffKJS2013}.
\end{center} \vspace{-0.4cm}
\end{table*}

Nowadays, an increasing number of sources have been reported in which spin-down rates undergo switches between two or more different states \citep{LyneHKS2010, KerrHJS2016, BrookKJK2016}. However, only a handful of pulsars showcase the correlations between spin behavior and alterations in pulsar radiation \citep{ChukwudeB2012, BrookKBR2014, TakataWLH2020}.
Some studies have explored the potential connection between glitches and changes in pulsar emission.
This link was first observed in PSR J1119$-$6227, where alterations in glitch behavior were found to influence the pulse profile \citep{WeltevredeJE2011}.
Subsequently, \cite{KeithSJ2013} conducted an in-depth investigation into the correlation between the spin-down rate of PSR J0742$-$2822 and its pulse shape, and found that there was a strong correlation after the occurrence of MJD $\sim$ 55022 glitch, suggesting a connection between glitches and shifts in the pulsar's radiation state.
An intriguing instance is observed in the case of PSR J2037+3621 (B2035+36), where a glitch at MJD $\sim$ 52950 led to a transition from a single radiation mode before the glitch to two distinct radiation modes (wide and narrow) after the glitch \citep{KouYWY2018}.
Further studies unveiled significant alterations in the average pulse profile width of PSR J2022+5154 (B2021+51), with a noticeable decrease before the glitch and a substantial increase following it \citep{LiuYST2021}. 
In a recent investigation, the post-glitch behavior of PSR J0738$-$4042 revealed a weakening of the leading peak in the integrated normalized pulse profiles, accompanied by an enhancement of the central component and a slight increase in pulse profile width \citep{ZhouGYG2023}\footnote{In a recent Bayesian analysis of the observed data, \citet{LowerJKB2023} obtained constraints $\Delta\nu/\nu\lesssim 2.6\times10^{-11}$ and $\Delta\dot\nu/\dot\nu=9(5)\times10^{-3}$ around the date of the pulse profile change for PSR J0738--4042 and disfavoured occurrence of a standard glitch for this pulsar. However, the scenario considered in \citet{ZhouGYG2023} involves both inward and outward motion of the superfluid vortex lines during the glitch driven by a crustquake, thus differing from a standard glitch in which only vortices migrating outward take part in the event. It should be noted that when almost equal number of vortices move inward and outward across a glitch, the change in the rotation rate would be tiny, while its effect stands out in the spin-down rate as this process alters the internal superfluid torque acting on the neutron star significantly \citep{GugercinogluA2020}. The observed temporal behavior also supports this point of view.  \citet{LowerJKB2023} have not taken this possibility into account.}.
Likewise, correlations between glitches and radiation changes have also 
been observed in the gamma-ray band.
For instance, following a glitch event occurring around MJD $\sim$ 55850, PSR J2021+4026 displayed a distinctive phenomenon where the third peak in the bridge region of the pulse profile vanished, leaving only two prominent peaks \citep{AllafortBBB2013, NgTC2016, ZhaoNLT2017, TakataWLH2020}. 
The most interesting finding was the abrupt 20\% decrease in post-glitch flux, rendering it the only gamma-ray pulsar known to exhibit such a pronounced flux jump \citep{AllafortBBB2013}.
In a separate investigation, \cite{LinWHT2021} analyzed 11-year timing data from the Fermi Large Area Telescope (Fermi-LAT) to identify four glitch events in PSR J1420$-$6048, and observed pulse profile changes following glitch events, underscoring the intricate interplay between glitch occurrences and radiation patterns.

For magnetars, the association between glitches and changes in radiation is more apparent.
Theoretically, magnetar glitches share similarities with rotation-powered radio pulsar glitches. 
For example, the long-term glitch activity of magnetars and radio pulsars are both proportional to their spin down rate~\citep{2017A&A...608A.131F}, implying a similar superfluid reservoir (core superfluid may be involved in large magnetar glitches~\citep{2008ApJ...673.1044D}. 
Besides, \citet{2014MNRAS.438L..16H} showed that glitch relaxations in anomalous X-ray pulsars could be naturally accounted for under the same framework as radio pulsars.
While from the view of observations, magnetar glitches manifested differently from radio pulsar glitches, about $20\%-30\%$ of magnetar glitches were associated with outbursts and violent pulse profile changes~\citep{2014ApJ...784...37D,2019AN....340..340H}, which are possibly the consequences of magnetic energy release following magnetar crustal fractures~\citep{2019MNRAS.488.5887S} or magnetic stress-induced magnetic field reconfiguration~\citep{RudermanZC1998}.

The present study focuses on PSR J1048$-$5832, an exceptionally unique pulsar exhibiting concurrent radiation emissions across three domains: radio, gamma-ray, and X-ray. 
Discovered through the Parkes high-frequency survey \citep{JohnstonSMK1992}, this pulsar was subsequently identified in X-rays by \cite{GonzalezKPG2006} and confirmed as a gamma-ray pulsar with relatively low significance by Fermi-LAT \citep{AbdoAAA2010}.
It is similar to a Vela-like pulsar due to its similarities with respect to rotational and radiation characteristics.
PSR J1048$-$5832 has a spin period of 123.7\,ms, a characteristic age of 20.4\,kyr, and a spin-down luminosity of $\dot{E} \sim 2.0 \times 10^{36}~{\rm erg~s^{-1} }$~\citep{JohnstonSMK1992,JankowskiBSK2019}. 
Its spin parameters and position are collected in Table \ref{Tab:1}.
Over its observational history, PSR J1048$-$5832 frequently glitches and has experienced 9 glitch events, with glitch sizes ($\Delta \nu / \nu$) ranging from $8.89(9) \times 10^{-9}$ to $3044.1(9) \times 10^{-9}$ \citep{WangMPB2000, Urama2002,YuMHJ2013,WeltevredeJMB2010, LowerDSR2021,ZubietaPGA2023, ZubietaMFL2023}. 
Among them, during the recovery process after five large glitches, not only did they show a long-term linear recovery, but most of them also included an exponential recovery process of dozens of days or even longer. 
\cite{YanMWW2020} explored the polarization properties, and identified the phase-stationary periodic amplitude modulation of this pulsar.

We analyze in this work the timing observation data of both Fermi-LAT and Parkes 64-m radio telescope from 2007 to 2023 for PSR J1048$-$5832.
The data span of Fermi-LAT and Parkes are displayed in the last two columns of Table \ref{Tab:1}.
Leveraging 16 years of comprehensive timing data, the present analysis successfully detects 5 glitch events. 
Importantly, our analysis reveals that the occurrence of glitches precipitated significant changes in the integrated mean pulse profile within the radio frequency domain. 
We also provide insights into the internal dynamics and interactions within neutron stars, as well as the complex interplay between the superfluid interior, crust, and magnetic field. 

The structure of this paper is as follows: 
In Section \ref{observations}, we briefly introduce the timing observations of Fermi-LAT and Parkes 64-m radio telescope and give an overview of the glitch data processing procedure. Results are presented in Section \ref{result}. Detailed discussions are made in Section \ref{discussion}, and a conclusion is provided in Section \ref{Summary}.
\begin{table*}
\caption{Pre- and post-glitch timing solutions for PSR J1048$-$5832: 
The first column lists the intervals corresponding to the glitch numbers. The second column is the time at which the spin parameters are measured. The third to fifth columns detail the spin parameters used in fitting the pulsar spin model, including frequency and its first and second derivatives. The next three columns describe the fitted data span, duration, and the number of ToAs within that span. The final column presents the root mean square (RMS) residual.
}
\label{Tab:F0F1-works}
\vspace{-0.3cm}
\begin{center}
    \renewcommand{\arraystretch}{1.4}
    \setlength{\tabcolsep}{11pt}      
\begin{tabular}{ccccccccc}
  \hline
    \hline       
Int.  & Epoch  & $\nu$  & $\dot{\nu}$ & $\ddot{\nu}$  & $N_{\rm ToA}$  & T  & MJD Range   & RMS residual\\
      & (MJD)  & (Hz)   & ($10^{-14}$ s$^{-2}$)   &  ($10^{-24}$ s$^{-3}$) & & (yr) & (MJD) & ($\mu$s) \\
    \hline     
5 -- 6   &54394  &8.0841086132(2)  &$-$625.959(2)   &$-$328(16)  &14   &0.5  &54302 -- 54486  &51    \\  
6 -- 7   &55625  &8.0834652518(9)  &$-$627.497(2)   &138(1)      &147  &6.1  &54504 -- 56747  &30301 \\  
7 -- 8   &56871  &8.082819068(1)   &$-$628.46(2)    &122(122)    &18   &0.6  &56766 -- 56977  &994   \\  
8 -- 9   &57997  &8.082203400(8)   &$-$627.136(1)   &140.9(8)    &119  &6.0  &56993 -- 59197  &22070 \\  
9 -- 10  &59379  &8.081455495(7)   &$-$625.81(1)    &98(63)      &11   &0.8  &59228 -- 59532  &1479  \\  
10 --    &59766  &8.081246304(1)   &$-$625.710(8)   &$-$49(25)   &15   &1.2  &59556 -- 59977  &1892  \\       
\hline     \hline  
\end{tabular}
\end{center} \vspace{-0.4cm}
\end{table*}


\section{Observations and Analysis}{\label{observations}}

\subsection{Fermi-LAT data}

LAT is the main instrument on board the Fermi satellite with an energy range of 20\,MeV -- 300\,GeV and has a large field of view of 2.4sr~\citep{AbdoAAA2010}, which is suitable to monitor the spin evolution of gamma-ray pulsar together with its survey mode. 
LAT is currently the most advanced high-energy gamma-ray telescope, boasting an effective area of 9500 $\rm cm^2$ \citep{AtwoodAAA2009}. It has excellent observational efficiency, completing an entire sky survey every two orbits (3 hr)~\citep{AbdoAAA2010}.
With a time accuracy better than 0.3\,$\mu{\rm s}$, LAT can easily detect even millisecond pulsars~\citep{AbdoAAA2010}. 
To date, Fermi-LAT continues to observe hundreds of gamma-ray pulsars, providing a unique opportunity to study their spin evolution.
LAT’s observation data of PSR J1048$-$5832 were collected from 2008 to 2022, covering a span of more than 14 years.
The timing analysis was performed using Fermi Science Tools software packages (v11r5p3) \footnote{\href{https://fermi.gsfc.nasa.gov/ssc/data/analysis/scitools/l}{https://fermi.gsfc.nasa.gov/ssc/data/analysis/scitools/}}.
The events were first filtered using \textit{gtselect} with an angular distance less than 0.5$^{\circ}$, zenith angle less than 105$^{\circ}$, and energy range 0.1 -- 10 GeV. Then the arrival time of each event was corrected to Solar system barycentre (SSB) using \textit{gtbary}. 
Each ToA was accumulated with an exposure time of at least 10 d for PSR J1048$-$5832, resulting in a total of 168 ToAs, as described by \cite{GeLYW2019}.

 \begin{table*}
\caption{The fitted timing solutions and glitch parameters for PSR J1048$-$5832. 
The first three rows present the frequency parameters of the fitted pulsar spin-down model, which align with the corresponding parameters in Table \ref{Tab:F0F1-works}. The fourth row indicates the reference epoch, a designated point in time slightly later than the glitch epoch. The fifth row specifies the glitch epoch. The subsequent three rows detail the data span, the number of ToAs, and the RMS, respectively. The final six rows in the table display the fitted glitch parameters.
The timing solution uncertainties are expressed as 1$\sigma$ values obtained through \texttt{TEMPO2}.} \vspace{-0.2cm}
\label{glitch:parameter}
\begin{center}
    \renewcommand{\arraystretch}{1.4}
    \setlength{\tabcolsep}{16pt}      
\begin{tabular}{lcccccc}
\hline
\hline  
Parameter             & Glitch 6         & Glitch 7        & Glitch 8        & Glitch 9        & Glitch 10    \\
\hline
$\nu$ (Hz)            & 8.084042089(4)  & 8.08282886(1)  &  8.082744132(6)  & 8.08151164(1)   & 8.081344107(5) \\
$\dot{\nu}$ ($10^{-12}$\rm\ s$^{-2}$)  
                      & $-$6.2597(4)    & $-$6.267(1)    & $-$6.286(1)   & $-$6.263(1)    & $-$6.2587(5)\\
$\ddot{\nu}$ ($10^{-24}$\rm\ s$^{-3}$)   
    & $-$23(27)    &$-$70(59)      & $-$218(127)   & $-$105(60)   & $-$34(25)\\ 
Freq. epoch (MJD)     & 54517           & 56800          & 57000  & 59275 & 59585\\
Glitch epoch (MJD)    & 54495(9)        & 56756(10)       & 56985(9)        & 59212(16)       & 59544(12) \\
Data span (MJD)       & 54301 -- 54860  & 56389 -- 56976 &  56795 -- 57168  & 59024 -- 59537  & 59219 -- 59982\\
ToA number           & 38              & 44              & 6              & 22            & 26 \\
RMS residual ($\mu {\rm s}$)     
                      & 948             & 3276            & 851            & 1455           & 1907          \\
$\Delta\nu$ ($10^{-6}$\rm\ Hz)   
                      & 24.594(4)       &  23.97(5)        &  0.022(3)       & 0.069(7)        & 0.035(4)\\
$\Delta\nu/\nu$ ($10^{-9}$)      
                      & 3042.4(5)       & 2965(6)         &  2.8(4)          & 8.6(8)         &  4.3(4) \\
$\Delta\dot{\nu}$ ($10^{-14}$\rm\ s$^{-2}$)     
                      & $-$2.73(8)      & $-$2.8(13)       &  0.3(2)    &  0.4(2)      & 0.22(8) \\
$\Delta\dot{\nu}/\dot{\nu}$ ($10^{-3}$)         
                      & 4.4(1)          & 4.5(22)          &  $-$0.5(3)         &  $-$0.6(2)     & $-$0.3(1) \\
$\tau_{\rm d}$            &   --             & 73(56)         &  --             & --              & --\\
$Q$                   &   --             & 0.004(2)        &  --             & --              & --\\
\hline     \hline  
\end{tabular}
\end{center} \vspace{-0.4cm}
\end{table*}

\subsection{Parkes data}

\begin{figure*}
\centering
\includegraphics[width=0.98\textwidth]{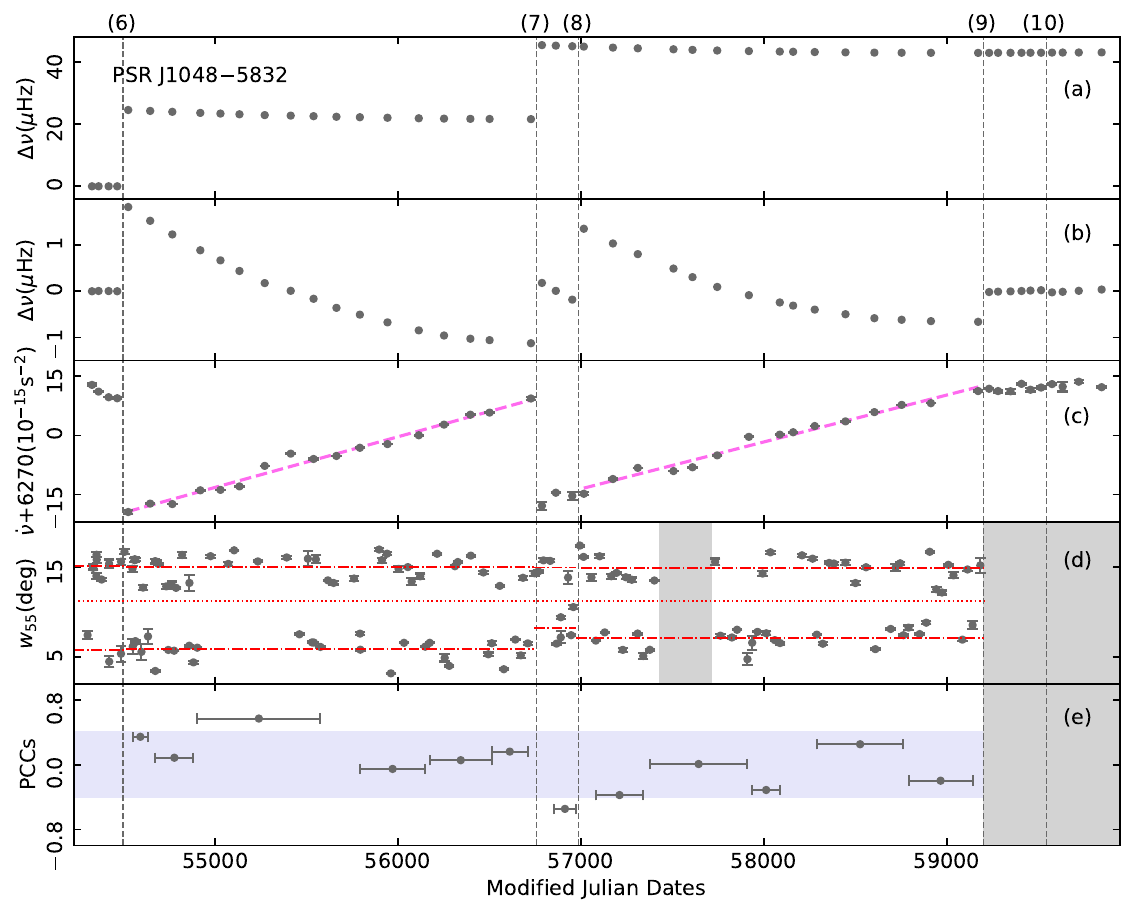}
\vspace{-0.4cm}
\caption{Glitches in PSR J1048$-$5832: panel (a) show pre- and post-glitch frequency residual ($\Delta \nu$) changes; panel (b) an expanded plot of $\Delta \nu$, represent the frequency residual of post-glitch relative to its mean value evolution; panel (c) the first frequency derivative ($\dot \nu$) evolves with time; panel (d) shows the $W_{\rm 55}$ of pulse profile; panel (e) is the Pearson cross-correlation coefficients (PCCs) between $\dot \nu$ and $W_{\rm 55}$; Vertical dashed lines indicate the glitch epochs. The numbers in parentheses are the pulsar glitch numbers; The red horizontal dash-dot line represents the average value of $W_{\rm 55}$ in both width and narrow modes; The red horizontal dot line is $W_{\rm 55} = 11^{\circ}$; The gray area and light blue area are the blank areas of the Parks data and the PCCs is $-$0.4 to 0.4, respectively.
}
\label{1048glitch}
\end{figure*}
\begin{figure}
\centering
\includegraphics[width=0.45\textwidth]{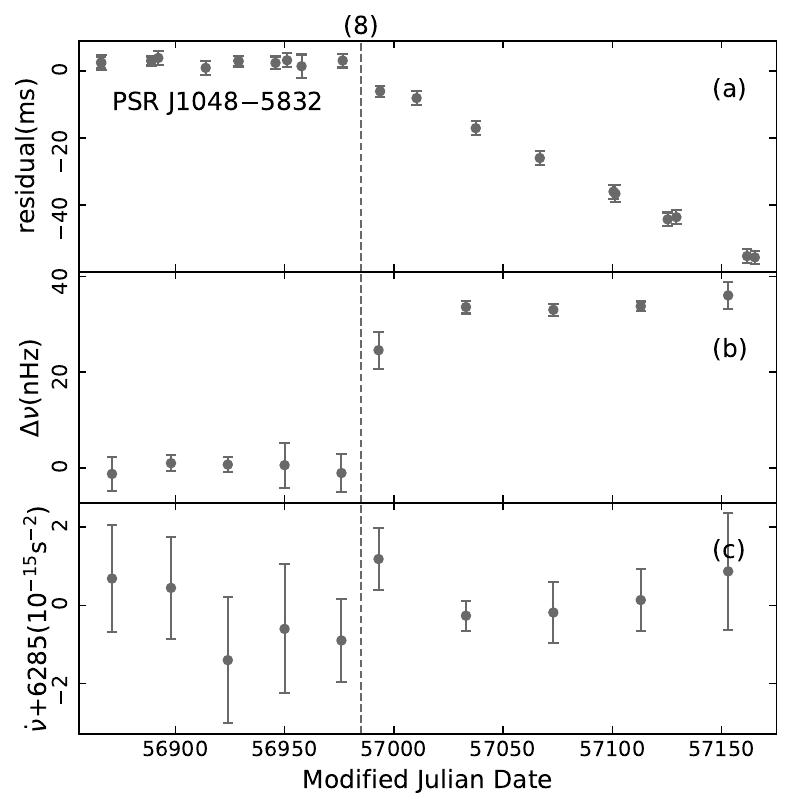}
\caption{Glitch 8 of PSR J1048$-$5832: Panel (a) shows the timing residual relative to the pre-glitch timing solution; Panel (b) is the spin frequency residual ($\Delta \nu$) relative to the pre-glitch timing solution; Panel (c) is the evolution of the first frequency derivative ($\dot \nu$). 
The vertical dashed line represents the glitch epoch; The number in the top parentheses is the pulsar glitch sequence.
} 
\label{glitch:8}
\end{figure}

We downloaded the timing data from PSR J1048$-$5832 at the online domain of Parkes pulsar data archive\footnote{\href{https://data.csiro.au/domain/atnf}{https://data.csiro.au/domain/atnf}}.
\cite{ManchesterHBC2013} introduced the timing observations of the Parkes radio telescope in detail. 
Briefly, Parkes uses a multi-beam receiver with a working wavelength of 20 cm for timing observations, with a centre frequency of 1369 MHz and bandwidth of 256 MHz \citep{StaveleyWBD1996,HobbsMDJ2020}.
Data was recorded using the Parkes Digital FilterBank system (PDFB1, PDFB2, PDFB3, and PDFB4).
The intervals of observations for PSR J1048$-$5832 are 2 -- 4 weeks with integration times of 2 -- 15 mins and sub-integration time of 30 s.

We utilized the pulsar data preprocessing and high-precision timing analysis software packages \texttt{PSRCHIVE} \citep{HotanVM2004, StratenDO2012} and \texttt{TEMPO2} \citep{HobbsEM2006,EdwardsHM2006} to analyze the Parkes timing data.
\texttt{PSRCHIVE} can be applied to eliminate the dispersion effect caused by the interstellar medium (ISM) and fold the sub-integration pulse profile to obtain the mean pulse profile.
In order to obtain more accurate pulse ToAs, we first updated the file header with ephemeris to align the phases of all mean pulse profiles. 
Then we used \texttt{PSRADD} to add these profiles to obtain a high signal-to-noise ratio standard pulse profile. 
Finally, we invoked the \texttt{PAT} tool of \texttt{PSRCHIVE} to cross-correlate each profile with the standard profile \citep{Taylor1992}.
Next, we used \texttt{TEMPO2} and Jet Propulsion Laboratory (JPL) DE430 Solar system ephemeris \citep{FolknerWBP2014} to transform these local ToAs into SSB and solve the timing solution.

With combining Fermi-LAT and Parkes timing observation data effectively, we used \texttt{TEMPO2} to achieve a series of spin frequencies and spin frequency derivatives.
It was noticed that the ToA errors of the two telescopes differ greatly, which may affect the timing solution.
Therefore, the \texttt{EFACEQUAD} plug-in was employed to obtain the parameters EFAC and EQUAD, adjusting the uncertainty of each ToA obtained from different terminals and telescopes.
EFAC corrects instrumental errors, while EQUAD accounts for an additional white noise process, as detailed in \cite{LentatiAHF2014}.
These two parameters were used when we solved the spin frequencies and spin frequency derivatives plotted in panels (a) and (b) of Fig. 1, and their acquisition did not involve Bayesian techniques.
After scaling the ToA errors by EFAC and EQUAD, the relationship between the new uncertainty (${\delta}{\rm s}$) and the initial uncertainty ($\delta$) is given by: ${\delta}{\rm s}^{2} = {\rm EFAC}^{2} \times (\delta^{2} + {\rm EQUAD}^{2})$.
After adding the error correction of ToA, the reduced chi-square values of each fitting region are distributed between 0.95 -- 1.05.
Following these adjustments, we utilized \texttt{TEMPO2} to fit these ToAs and obtain the best rotation models for this pulsar.  
We subsequently calculated timing residuals by subtracting the ToAs predicted by this model from the actual observed ToAs.

According to the rotation model, the evolution of the pulsar rotation phase $\phi(t) $ can be given by Taylor series expansion \citep{EdwardsHM2006}: 
\begin{equation}
\label{equ:1} 
\phi(t) = \phi(t_{\rm 0}) + \nu(t - t_{\rm 0}) + \frac{\dot{\nu}}{2} (t - t_{\rm 0})^{2} + \frac{\ddot{\nu}}{6}(t - t_{\rm 0})^{3} \ ,
\end{equation}
where $\nu$, $\dot{\nu}$, $\ddot{\nu}$ represent the rotation frequency and its first and second-time derivatives, respectively. 
$\phi(t_{\rm 0})$ is the pulse phase at a reference epoch $t_{\rm 0}$, and it can be assumed that $\phi(t_{\rm 0}) = 0$ at $t = t_{\rm 0}$.
During a glitch, a pulsar's rotation frequency can experience a rapid increase, followed by a gradual recovery to its original rate. 
This recovery process is typically exponential in nature followed by a linear relaxation of the increase in the spin-down rate with a constant second time derivative of the spin frequency.
Therefore, we can use the change of $\nu$, $\dot{\nu}$, and the possible exponential recovery process to express the post-glitch pulse phase change \citep{EdwardsHM2006,YuMHJ2013}: 
\begin{equation}
\begin{split}
\label{equ:2} 
\phi_{\rm g} = \Delta\phi+ \Delta\nu_{\rm p}(t - t_{\rm g}) +   \frac{1} {2} \Delta\dot{\nu}_{\rm p} (t - t_{\rm g})^{2} \\
+ \sum_{\rm i} {\Delta\nu_{\rm d_{i}}\tau_{\rm d_{i}} [1-e^{-(t - t_{\rm g})/\tau_{\rm d_{\rm i}}}] }  \ ,
\end{split}
\end{equation}
where $\Delta\phi$ is offset of pulse phase, $\Delta\nu_{\rm p}$ and $\Delta\dot{\nu}_{\rm p}$ are the permanent increments of the pulsar rotation frequency and its first derivative, respectively.
$t_{\rm g}$ represents the epoch at which the glitch occurs.
$\Delta\nu_{\rm d_{i}}$ refers to the recovery range of the rotation frequency after $\tau_{\rm d_{i}}$ time during the post-glitch recovery process.
So we can rewrite the fractional glitch size as:   
\begin{equation}
\label{equ:3} 
 \frac{\Delta\nu} {\nu} = \frac{\Delta\nu_{\rm p} + \sum_{\rm i} {\Delta\nu_{d_{i}}}} {\nu} \ ,
\end{equation}

\begin{equation}
\label{equ:4} 
 \frac{\Delta\dot{\nu}} {\dot{\nu}} = \frac{\Delta\dot{\nu}_{\rm p} - \sum_{\rm i} {\Delta\nu_{\rm d_{i}} / \tau_{\rm d_{i}}}} {\dot{\nu}} \ .
\end{equation}
The glitch recovery factor $Q$ is defined as:
$\frac{ \sum_{\rm i} {\Delta\nu_{\rm d_{i}}} } {\sum_{\rm i} {\Delta\nu_{\rm d_{i}}} + \Delta\nu_{\rm p}}$.

\section{Result}\label{result} 

\subsection{Timing solutions and glitch parameters}

By combining the ToAs derived from both Fermi-LAT and Parkes observations, we conducted a comprehensive analysis of the spin evolution of PSR J1048$-$5832 spanning nearly 16 years, from 2007 to 2023.
In our timing data analysis, we identify a total of five glitch events, labeled as glitches 6, 7, 8, 9, and 10. 
Table \ref{Tab:F0F1-works} presents the pre- and post-glitch timing solutions.
The glitch epoch is determined as the central date between the last pre-glitch observation and the first post-glitch observation, with an uncertainty equal to half the observation interval.
The reported parameter uncertainties are presented as 1$\sigma$ values derived from \texttt{TEMPO2}.
We update the timing solutions and detailed parameters of glitches 6, 7, 8, 9 and 10, as documented in Table \ref{glitch:parameter}.
The glitch parameters tabulated in Table \ref{glitch:parameter} are determined by fitting Equation (\ref{equ:2}), with their associated uncertainties derived using the error transfer equation.
Figure \ref{1048glitch}, panels (a), (b), and (c), illustrate the evolution of spin frequency and its first derivative for PSR J1048$-$5832 under the influence of glitches.
Panels (a) and (c) exhibit good consistency with the glitch parameters detailed in Table \ref{glitch:parameter}.

Among the five glitch events, glitch 8 (MJD $\sim$ 56985 (9)) represents a new small event detected more than 200 d following glitch 7. 
The size of the glitch is only $\Delta \nu/\nu \sim  2.8(4)\times10^{-9}$, and the relative change in spin-down rate is $\Delta \dot{\nu}/\dot{\nu} \sim  -5(3)\times10^{-4}$, which is currently the smallest glitch detected from this particular source to date.
As shown in Figure \ref{glitch:8}, panel (a) illustrates the timing residuals relative to the spin-down model prior to the glitch.
This conforms to the typical pattern observed in small glitches, after which a significant linear deflection is evident.
In panels (b) and (c), we observe a noticeable discontinuity in both $\Delta \nu$ and $\dot{\nu}$, further confirming the occurrence of this small glitch.
Due to the absence of significant exponential recovery features in (b) and (c), we did not include an exponential term in the glitch parameter fitting.
For other parameters of glitch 8, please refer to Tables \ref{Tab:F0F1-works} and \ref{glitch:parameter}.

\begin{figure*}
    \centering
    \includegraphics[width=1\linewidth]{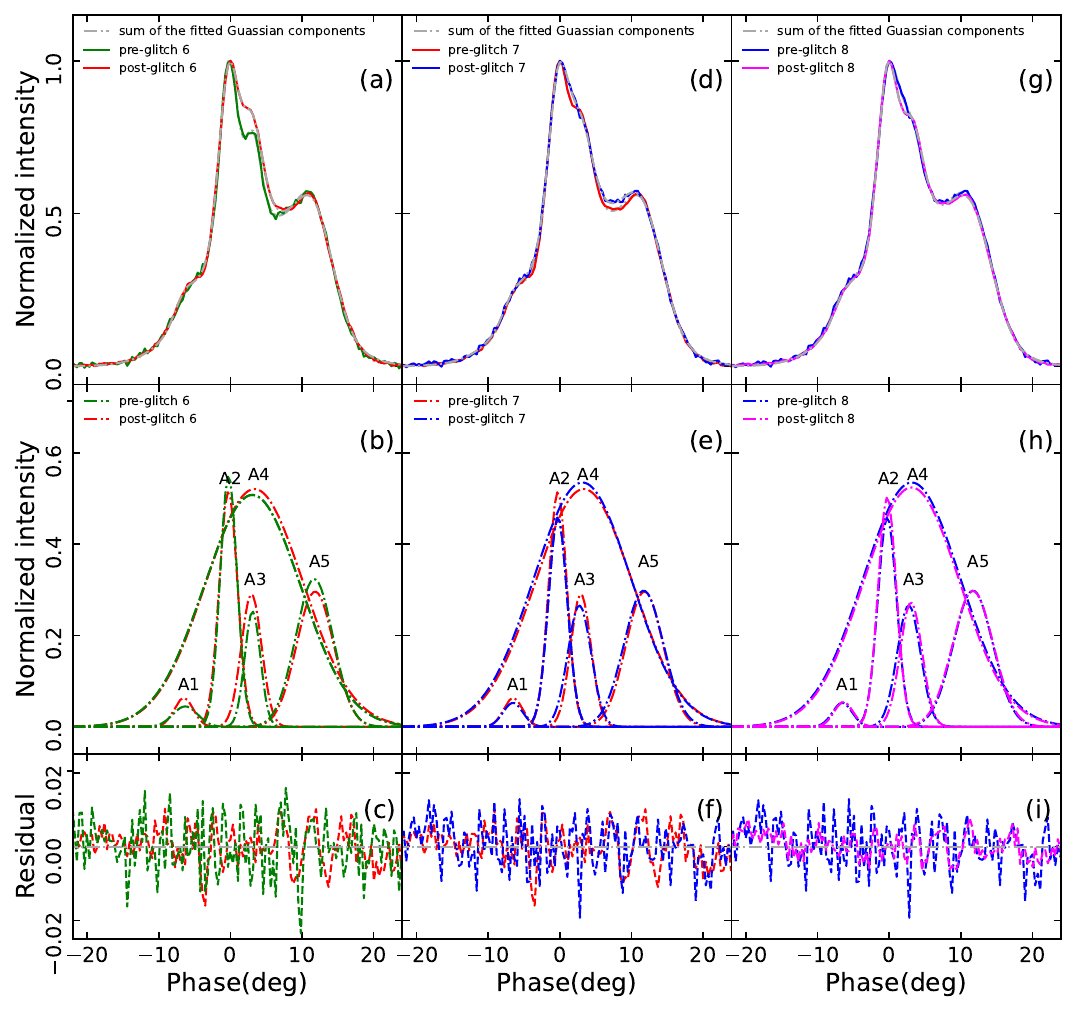}
\vspace{-0.4cm}
    \caption{The integrated normalized pulse profile and profile fitting. 
 Panels (a), (d) and (g) are the pulse profiles before and after the glitch, respectively, and the gray lines are the pulse profile of fitting. 
 In panel (a), 
 the green curve represents the integrated and normalized pulse profiles observed prior to glitch 6. This dataset consists of 11 observations with high signal-to-noise ratio in the range of MJD 54302 -- 54486, with an integration time of 30 minutes total;
The red curve depicts the integrated, normalized pulse profiles following glitch 6, encompassing 68 observations between MJD 54504 -- 56747, with a total integration time of 3.1 hours.
In panel (d), the blue curve illustrates the mean pulse profiles observed after glitch 7 (MJD 56766 -- 56977), with a combined integration time of 29 minutes.
In panel (g), the magenta curve illustrates the mean pulse profiles observed after glitch 8 (MJD 56993 -- 59197), with a combined integration time of 2.6 hours.
Panels (b), (e) and (h) show the five Gaussian components that fit the pulse profile. 
The green, red, blue and magenta curves in the figure represent pre-glitch 6, post-glitch 6, post-glitch 7, and post-glitch 8, respectively. 
Panels (c), (f) and (i) show the residuals of the fitted profile and the corresponding pulse profile.}
\label{1048profile}
\end{figure*}

The timing solutions and glitch parameters presented in Tables \ref{Tab:F0F1-works} and \ref{glitch:parameter} are generally consistent with the previously reported results~\citep{WeltevredeJMB2010,YuMHJ2013,LowerDSR2021,ZubietaPGA2023,ZubietaMFL2023}, although there are some discrepancies. 
For glitch 6, its glitch parameters are basically consistent with the findings of \citet{WeltevredeJMB2010}.
In the case of glitch 7, there are some differences from the results reported by \citet{LowerDSR2021}. 
Specifically, the relative change in spin-down rate $\Delta \dot{\nu}/\dot{\nu}$ is only half of the previously reported values. 
Glitches 9 and 10 are both small glitch events, as previously reported by \citet{ZubietaPGA2023,ZubietaMFL2023}. 
For glitch 9, our fitting determined that the glitch size is $\Delta \nu \sim 6.9(7) \times 10^{-8}$ Hz, which is consistent with the reported value ($\Delta \nu \sim 7.19(7) \times 10^{-8}$ Hz) by \cite{ZubietaPGA2023,ZubietaMFL2023} within the error range.
Concerning glitch 10, our determined glitch size ($\Delta \nu \sim 3.5(4) \times 10^{-8}$ Hz) is less than half of the previously reported value ($\Delta \nu \sim 8.02(25) \times 10^{-8}$ Hz)~\citep{ZubietaPGA2023,ZubietaMFL2023}. 

In contrast to the standard glitches, the fractional changes in the spin frequency derivative following glitches 8, 9 and 10 are, though small, all negative, i.e. $\Delta\dot\nu/\dot\nu<0$ (see Table \ref{glitch:parameter}). Moreover, as can be easily recognized from Figures \ref{1048glitch} and \ref{glitch:8} (also refer to Figures 3 and 4 in \citet{ZubietaPGA2023}), after each of these glitches, the spin frequency derivative exhibits low-amplitude oscillations. The underlying reason for both phenomena may be the time-variable internal superfluid torque exerted on the neutron star's crust \citep{erbil2023}, in response to the changes in the magnetospheric state resulting in the periodic modulation observed by \citet{YanMWW2020}.

We note that, following glitches 6 and 8, the evolution of $\dot{\nu}$ seems to display a linear trend dominated by inter-glitch relaxation. 
This manifests as a relatively stable spin frequency second derivative ($\ddot{\nu}$), which can be clearly seen in panel (c) of Figure \ref{1048glitch}.
Moreover, Table \ref{Tab:F0F1-works} indicates that both of these glitches exhibit an inter-glitch interval of up to 6 years. In light of this, we 
calculate the pulsar braking index ($n = \nu \ddot{\nu}/\dot{\nu}^2$).
We use the least squares method to directly perform linear fitting on the slope of $\dot{\nu}$ corresponding to the recovery phases of glitches 6 and 8.
The resulting $\ddot{\nu}$ values are $1.486(10) \times 10^{-22} ~{\rm s}^{-2}$ and $1.379(8) \times 10^{-22} ~{\rm s}^{-2}$ for glitches 6 and 8, respectively.
The fitting results are represented by the bright magenta dashed lines in panel (c). 
We further extract the values of $\nu$ and $\dot{\nu}$ within the post-glitch 6 and 8 periods, as detailed in Table \ref{Tab:F0F1-works}. By fitting these corresponding $\ddot{\nu}$ values, we calculate the braking index to be 30.5(2) and 28.3(2) 
for post-glitches 6 and 8, respectively.
These braking index values are generally consistent with the results of \citet{LowerDSR2021} and can be explained through pulsar acceleration due to decoupling of part of the crustal superfluid from the pulsar braking torque during the glitch \citep{AlparB2006}.

 \begin{table*}
\caption{Pulse width ($W_{\rm 55}$) of each mean pulse profile and pulse equivalent width ($w_{\rm eq}$) of the integrated mean pulse profiles in different periods. 
} \label{profile:parameter} 
    \renewcommand{\arraystretch}{1.15}
    \setlength{\tabcolsep}{18pt} 
 \begin{tabular}{ccccc}
\hline \hline
Parameter  & Pre-glitch 6 & Post-glitch 6  &Post-glitch 7 & Post-glitch 8 \\
\hline
Data span (MJD)           &54302 -- 54486   &54504 -- 56747 &56766 -- 56977  &56993 -- 59197\\
Pulse number                 &11            &68         &9     &53  \\
Integrated time (hours)      &0.5           &3.1        &0.5       &2.6 \\
Mean narrow $W_{\rm 55}$ (deg)   &5.7(3)           &5.8(1)     &8.2(1)       &7.07(5) \\
Mean wide $W_{\rm 55}$ (deg)     &15.1(1)          &14.9(1)    &14.9(2)     &14.90(5) \\
Integrated mean pulse profile $w_{\rm eq}$ (deg)       &13.23     &13.69     &13.96        &13.76\\
\hline \hline
\end{tabular}
\vspace{-0.5cm}
\end{table*}

\subsection{Variation of the radio pulse profile}
\label{sec:3.2}
In the radio band, we analyze the pulse profiles between MJD 54302 -- 59182.
Since no radio data is available following glitches 9 and 10, our analysis primarily centers on assessing changes in the pulse profile before and after glitches 6, 7 and 8.
Building upon the updated average pulse profiles detailed in the previous section, we undertake a phase alignment and superposition of all profiles within the corresponding temporal windows. This process yields the integrated mean pulse profiles, which are displayed in Figure \ref{1048profile}.
Panels (a), (d) and (g) display the variation of the integrated mean pulse profiles before and after glitches, respectively.
To ensure comparability, we normalize the integrated pulse profile by dividing the amplitude by its maximum value and also align the peak phase.
As illustrated in panel (a) of Figure \ref{1048profile}, the integrated mean profile of PSR J1048$-$5832 exhibits one main peak, one leading peak, and two trailing peaks.
For glitch 6, a noticeable variation in the integrated mean pulse profile is evident when comparing the pre- and post-glitch states.
Particularly noteworthy is the significant increase in the relative intensity of the first trailing peak component observed in the post-glitch period. 
While the figure may not directly convey a substantial change in pulse width, calculations reveal a slight increase in the pulse equivalent width when comparing the pre- and post-glitch states. Detailed values of pulse widths can be found in Table~\ref{profile:parameter}. 

\begin{table*} 
\caption
{The parameters of the Gaussian components of the integrated mean pulse profiles. 
}\label{gaussian:parameter}
    \renewcommand{\arraystretch}{1.15}
    \setlength{\tabcolsep}{6pt} 
 \begin{tabular}{ccccccccccccc}
\hline
\hline
\multicolumn{0}{c}{Gaussian component}& \multicolumn{3}{c}{ Pre-glitch 6 } & \multicolumn{3}{c}{Post-glitch 6} &  \multicolumn{3}{c}{Post-glitch 7} & \multicolumn{3}{c}{Post-glitch 8} \\
& $w_{\rm eq}$  (deg)  & $h$   & $p$ (deg) 
& $w_{\rm eq}$  (deg)  & $h$   & $p$ (deg) 
& $w_{\rm eq}$  (deg)  & $h$   & $p$ (deg) 
& $w_{\rm eq}$  (deg)  & $h$   & $p$ (deg)   \\
\hline
A1  &4.38   &0.04  &$-$6.31  &3.37   &0.06  &$-$6.52   &3.46  &0.05  &$-$6.55   &3.36  &0.05  &$-$6.39 \\  
A2  &2.97   &0.55  &$-$0.25  &3.21   &0.51  &$-$0.29   &3.32  &0.46  &$-$0.27   &3.22  &0.50  &$-$0.28\\ 
A3  &2.78   &0.25  &3.09     &3.61   &0.29  &2.93      &4.09  &0.27  &2.73      &3.85  &0.27  &3.05\\ 
A4  &16.92  &0.51  &3.06     &17.04  &0.52  &3.36      &17.03 &0.54  &3.15      &16.80 &0.52  &3.13\\ 
A5  &6.60   &0.32  &11.71    &6.53   &0.30  &11.83     &6.97  &0.30  &11.71     &7.17  &0.30  &11.70\\ 
\hline
\end{tabular} 
 \end{table*}

In order to better judge whether the integrated profile changes after the glitch, we select a section of the baseline with no radiation and high noise in the pulse profile to calculate the standard error ($\sigma$).
The difference between the pulse profiles of pre- and post-glitch is used as the profile residual.
When the absolute value of the profile residuals of at least three consecutive bins is greater than 3$\sigma$, the pulse profile within this longitude range is considered to have changed.
According to the method described above, it is confirmed that the integrated mean pulse profile of post-glitch 6 changes within the longitude range of $0.35^{\circ}$ to $6.33^{\circ}$.

Turning to panel (d) of Figure \ref{1048profile}, it becomes apparent that the integrated normalized mean profile is also different after glitch 7.
Firstly, the intensities of the leading peak and the two trailing peaks of the integrated pulse profile intensify following glitch 7.
Secondly, the profile residuals in the three longitude ranges of $-3.87^{\circ}$ to $-3.16^{\circ}$, $0.70^{\circ}$ to $2.81^{\circ}$, and $5.27^{\circ}$ to $9.49^{\circ}$ are all more than 3$\sigma$, indicating that the integrated pulse profile of post-glitch 7 has changed in these areas, which is consistent with the results obtained by directly observing panel (d).
While these changes in the pulse profile's shape are evident, the equivalent width of the integrated profile remains relatively stable following the glitches. Detailed parameters are provided in Table \ref{profile:parameter}.

For glitch 8, the integrated normalized mean pulse profiles of pre- and post-glitch are shown in panel (g) of Figure \ref{1048profile}.
From panel (g), it can be seen that compared to glitch 6 and 7, the changes in the integrated pulse profile before and after glitch 8 are relatively small.
The main manifestation is that the intensity of the first trailing peak of post-glitch 8 is relatively weakened, while the intensity of the second trailing peak also seems to be diminishing.
After calculation, it was confirmed that the integrated profile changed in the longitude range of the first trailing peak ($0.70^{\circ}$ -- $2.46^{\circ}$), while the profile residual of the second trailing peak was less than 3$\sigma$, thus ruling out the possibility of a change in this component.

In order to obtain detailed information about the shape changes of the pulse components at pre- and post-glitch,  we employ Gaussian components for fitting and distinguishing these features, as outlined in \citet{Kramer1994}.
Utilizing Gaussian components allows for precise determination of profile features, including the number, position, width, and flux of these components, as demonstrated in studies such as \cite{FergusonBWB1981} and \cite{KramerWJS1994}.
Analyzing the change in Gaussian components before and after a glitch proves to be valuable in examining variations in the pulsar's radiation region due to glitch.
For instance, in \cite{LiuYST2021}, it is postulated that each Gaussian component corresponds to a flux tube within the radiation region, which enables discussions regarding the motion of these flux tubes in response to the changes in Gaussian components. 
Subsequently, these authors reported similar changes in the average pulse profile of PSR J1825$-$0935 (B1822$-$09) following a glitch event, attributing these variations to modifications in the Gaussian components~\citep{LiuSYT2022}.
We here utilize the nonlinear least squares method to fit the integrated normalized mean pulse profiles. 
For each pulse profile, we conduct a total of 80,000 Gaussian fitting attempts. 
From the post-fit residual diagram in the lower panel, it can be seen that the residual exhibits minimal fluctuations, indicating that the fitting results can conform well to the integrated pulse profile.
Panels (b), (e) and (h) show the Gaussian component fitting process applied to the integrated normalized mean pulse profile of PSR J1048$-$5832, resulting in five Gaussian components denoted as A1, A2, A3, A4, and A5. 
Specific Gaussian component fitting parameters are detailed in Table  \ref{gaussian:parameter}, where  $w_{\rm eq}$, $h$, and $p$ correspond to the pulse effective width, the peak pulse intensity, and the phase at the peak intensity, respectively.
Panels (c), (f) and (i) in  Figure \ref{1048profile} show the residual distribution of the fitted pulse profile, further corroborating the compatibility of the fitting outcomes with the integrated mean pulse profile.  

From panels (b), (e) and (h) of Figure \ref{1048profile}, it can be directly observed that the Gaussian components of each integrated pulse profile are consistent, and the changes in the integrated pulse profile can correspond to the change of the Gaussian components.
For glitch 6, it can be intuitively observed from panel (b) that each Gaussian component undergoes a variation after the glitch, and they have different contributions to the integrated pulse profile.
These changes in components A3 and A4 contribute to an overall increase of pulse intensity within the longitude range of 0.35$^{\circ}$ to 6.33$^{\circ}$ in the integrated profile.
Moreover, the $h$ values for components A2 and A5 undergo reductions. 
Though the intensity of A4 increased, the second trailing peak of the integrated pulse profile remained unchanged under the influence of decreased A5.  
However, in light of the influence exerted by A1, A2 and A4, the leading and main peaks of the integral profile remains relatively stable despite these adjustments.

Panel (e) provides valuable insights into the differences between the integrated pulse profiles pre- and post-glitch 7, primarily driven by changes in components A4 and A5.
Comparing the pulse profiles before and after glitch 7, it is evident that the pulse effective width of component A5 has increased by $0.44^{\circ}$. This widening results in an increased intensity of the integrated pulse profile within the longitude range of $5.27^{\circ}$ -- $9.49^{\circ}$.
Furthermore, the integrated average profile shows an increase in pulse intensity within the longitude range of $-3.87^{\circ}$ -- $-3.16^{\circ}$ and $0.70^{\circ}$ -- $2.81^{\circ}$. 
This increase is primarily attributable to a 0.02 increase in the peak intensity of A4 after glitch 7, accompanied by a leftward shift of $0.21^{\circ}$ in the peak phase.
Additionally, after glitch 7, Gaussian components A1, A2, and A3 also exhibit some variations. However, due to the interplay among these Gaussian components, their influence on the overall change in the integrated mean pulse profile is relatively small.
Panel (h) shows the variations in the Gaussian component of pre- and post-glitch 8.
This figure shows that there is hardly any change in components A1 and A5 after the glitch.
Moreover, although the intensity of component A2 has increased, the integral pulse profile remains unchanged due to the weakening of component A3.
Finally, the pulse intensity of components A3 and A4 weaken, which causes the integrated profile intensity of post-glitch 8 to decrease in the longitude range from $0.70^{\circ}$ to $2.46^{\circ}$.

Furthermore, it is observed that the Gaussian component of the integrated normalized mean pulse profile changed differently after glitches 6, 7 and 8.
The differences suggest that glitches with varying jump amplitudes may exert distinct effects on the same Gaussian component. In other words, different glitches might impact the same magnetic flux tube within the radiation region differently, as discussed in \citet{LiuSYT2022}.
Given that the jump amplitude of glitch 6 substantially surpasses that of glitches 7 and 8, the changes in Gaussian components subsequent to glitch 6 are notably more pronounced. 
This supports the hypothesis that the integrated mean pulse profile may exhibit significant variability within a short timescale due to periodic mode change.
Indeed, \cite{YanMWW2020} reported that the central and trailing components of the pulse profile of PSR J1048$-$5832 intermittently switch between strong and weak modes. 
Consequently, we do not rule out the possibility that the changes in integrated mean pulse profiles before and after glitches 6, 7, and 8 may be influenced by mode changes.

In addition, we study the polarization properties of PSR J1048-5832 using Parkes data and analyze the polarization before and after glitches 6, 7, and 8, respectively. 
The results suggest that glitches do not significantly affect polarization. 
Similar conclusions were previously reported in \cite{YanMWW2020}.

\subsection{Correlation between the spin-down rate and radio emission}

To explore the relationship between the spin-down rate and radiation, we conduct measurements of the full width of each observation pulse profile at the 55\% peak amplitude ($W_{\rm 55}$) since its evolution with glitch is most pronounced. 
Panel (d) in Figure \ref{1048glitch}, illustrates the temporal evolution of $W_{\rm 55}$ in the Parkes data.
Visually, the figure effectively depicts variations in the average pulse profile width across both wide and narrow modes. 
For our analysis, we defined $W_{55} > 11^{\circ}$ as the wide mode and $W_{55} < 11^{\circ}$ as the narrow mode. 
The average value of $W_{55}$ in the wide and narrow modes of each period are detailed in Table~\ref{profile:parameter}, and are represented by the red horizontal dash-dot lines in panel (d) of Figure \ref{1048glitch}.
It is clear that within the wide mode, there is no substantial change in $W_{55}$, and the mean value for each period is approximately 15$^{\circ}$.
However, within the narrow mode, we observe noteworthy differences. For glitch 6, the post-glitch $W_{55}$ (5.7(3)$^{\circ}$) exhibits no significant change compared to the pre-glitch value of $W_{55}$ (5.8(1)$^{\circ}$). This could be attributed to the limited pre-glitch observations.
Following glitch 7, a clear increase in the $W_{55}$ within the narrow mode is evident when compared to the measurement after glitch 6. Specifically, the average value of $W_{55}$ after glitch 6, which stands at 5.8(1)$^{\circ}$, rises to 8.2(1)$^{\circ}$ after glitch 7.
For glitch 8, compared to pre-glitch, the average value of $W_{55}$ of post-glitch (7.07(5)$^{\circ}$) is significantly reduced, and the difference is $-1.1(2)^{\circ}$.

From panels (c) and (d) in Figure \ref{1048glitch}, it is not immediately evident whether a correlation exists between the pulsar's spin-down rate and the evolution of $W_{55}$.
To further explore their correlation, we examine the Pearson cross-correlation coefficients ($\alpha$) between $W_{55}$ in narrow modes and $\dot{\nu}$.
Considering the sparse observational data for the narrow mode preceding glitch 6, we exclude this data from our analysis and focus only on the data set nearly 13 years following glitches 6, 7 and 8. 
To obtain $\alpha$, we employ a linear interpolation method to match $\dot{\nu}$ with $W_{55}$.
We calculate a $\alpha$ value every 5 observations (about 84-670 d) and show the results in panel (e) of Figure \ref{1048glitch}.
The light blue area in the figure corresponds to the range $-0.4 \leq \alpha \leq 0.4$. When $\alpha$ falls within this range, it indicates a lack of significant correlation between $W_{55}$ and $\dot{\nu}$.
As observed in the figure, the distribution of $\alpha$ is scattered across the entire span, with the majority falling within the $-0.4 \leq \alpha \leq 0.4$ range. 
This confirms that no statistically significant correlation was detected between $W_{55}$ in the narrow mode and $\dot{\nu}$.
As elaborated above in Section \ref{sec:3.2}, despite the absence of a correlation between spin and radiation, the evolution of pulse profiles within narrow modes still suggests that changes in the integrated mean profile of PSR J1048$-$5832 could be linked to the glitch event.

\begin{figure}
\centering
\includegraphics[width=0.45\textwidth]{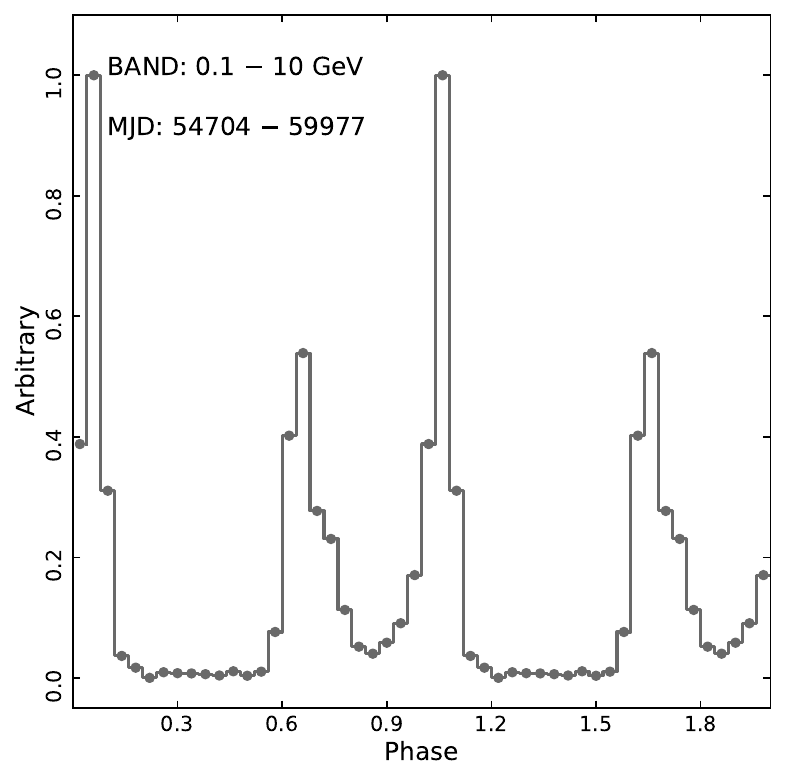}
\vspace{-0.3cm}
\caption{The normalized gamma-ray pulse profile of PSR J1048$-$5832 over the entire data span (0.1-10 GeV).}
\label{gama:profile}
\end{figure}

\begin{figure*}
\centering
\includegraphics[width=1\textwidth]{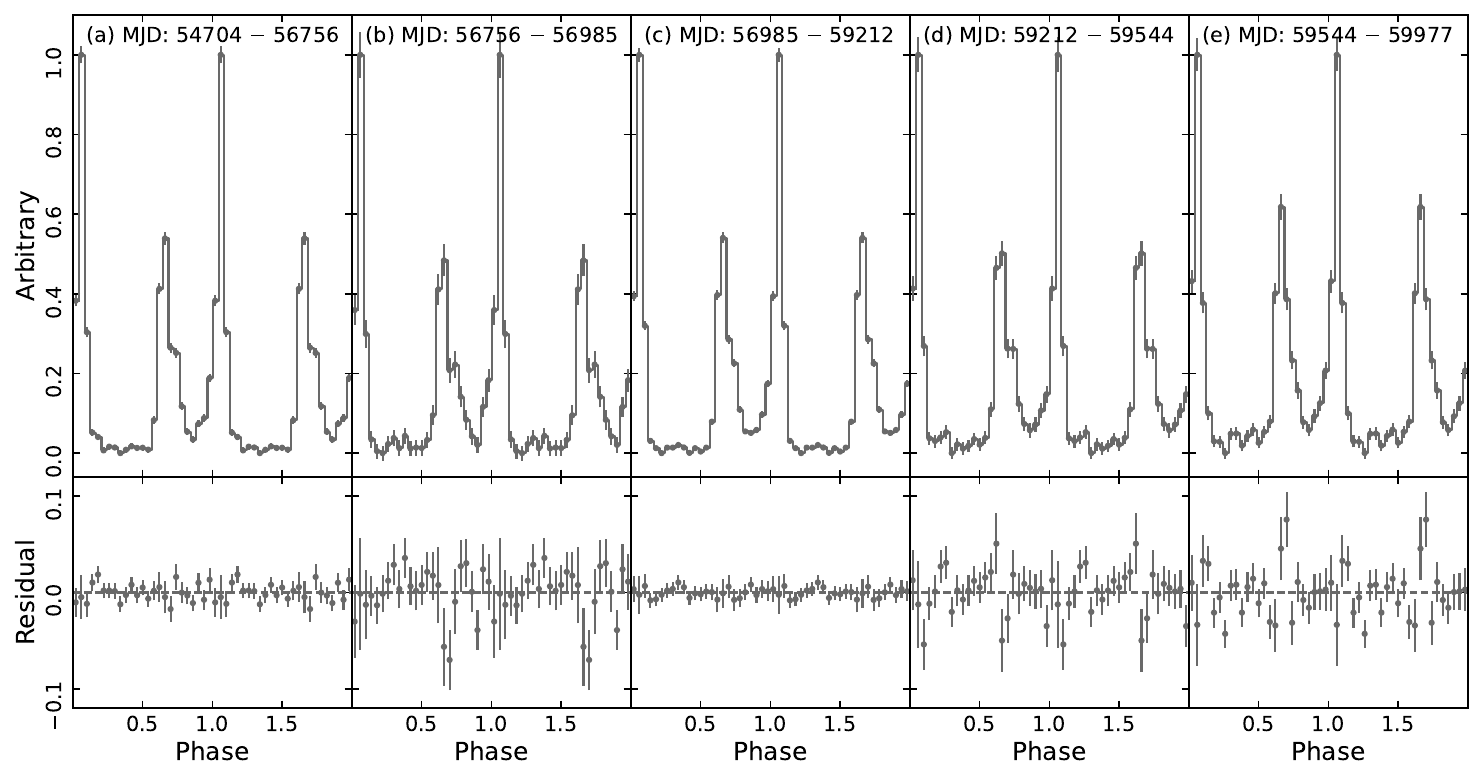}
\vspace{-0.5cm}
\caption{The normalized gamma-ray pulse profile of PSR J1048$-$5832 over the corresponding time span. Panels (a) -- (e) represent post-glitch for glitches 6, 7, 8, 9, and 10, respectively. The lower panel shows the differences between the profile of each panel and the total integral profile.}
\label{gama:glitch}
\end{figure*} 


\subsection{Variation of the gamma-ray pulse profile}

We examine the gamma-ray integrated pulse profile of PSR J1048$-$5832 both before and after the glitch event, as depicted in Figure \ref{gama:profile} and Figure \ref{gama:glitch}. 
Due to the limited number of photons with energies exceeding 10 GeV, we restrict our analysis to photons with energies ranging from 0.1 to 10 GeV to obtain the gamma-ray pulse profile.
Figure \ref{gama:profile} shows the total normalized integrated pulse profile of PSR J1048$-$5832 across the time spanning of MJD 54704 -- 59977 (0.1 -- 10 GeV).
 To further investigate the impact of the glitch on the gamma-ray pulse profile, we divide the data into five intervals, denoted as (a), (b), (c), (d), and (e), based on the glitch epoch. We then generate a pulse profile for each interval. The gamma-ray integrated pulse profiles for each interval are presented in Figure \ref{gama:glitch}, and the residuals between each profile and the total integrated profile are displayed in the lower panel.
From Figure \ref{gama:glitch}, it is evident that the contours of each interval exhibit varying degrees of variation. Panels (a) and (c) in Figure \ref{gama:glitch} reveal that the residuals of the pulse profile following glitches 6 and 8 are uniformly distributed, with very low Chi-square values (0.65 and 0.54, respectively, for 40 degrees of freedom (d.o.f.)). This indicates a close similarity between the pulse profiles in these cases.
Conversely, for panels (b), (d), and (e), substantial changes are observed in the contour residuals of these three intervals, accompanied by Chi-square values (40 d.o.f.) of 0.82, 1.26, and 1.98, respectively. Notably, the Chi-square values for the pulse profiles in intervals from (a) through (e) all remain below 3, indicating the absence of significant changes relative to the total integrated profile.
We attribute the smaller residuals in intervals (a) and (c) to their longer integration times, exceeding 6 years. Conversely, the profiles in intervals (b), (d), and (e) exhibit more substantial changes that may be attributed to statistical fluctuations.

\section{Discussion}
\label{discussion}

\subsection{Understanding the glitch effects on profile change}

The widely accepted glitch model is the superfluid model~\citep{1969Natur.224..872B,AndersonN1975}. 
The observational characteristics of the quite short spin-up timescale, relatively slow relaxation behavior, repetition times, and the lack of a significant increase in the surface temperature of neutron stars associated with the energy dissipating sudden jumps all indicate that the superfluid internal structure of neutron stars is responsible for the pulsar glitches [See the reviews \citep{HaskellM2015,ZhouGYGY2022,AntonopoulouHE2022,AntonelliMP2022} for superfluid aspects of neutron stars and their glitches]. 
The neutron star's internal superfluid behaves akin to a reservoir for angular momentum, which releases angular momentum and transfers it to the crust under certain conditions, thereby causing a sudden increase in the neutron star's spin frequency.

In order for the core superfluid to manage to spin down to keep up with the rotational angular velocity of the normal matter crust and in turn neutron star surface, the vortex lines within the neutron star core must migrate radially outward with speed a $\sim R_{*}/(4t_{\rm sd})$ with $R_{*}$ and $t_{\rm sd}$ being neutron star's radius and characteristic age, respectively. While traversing the core region, these vortices cut through magnetic flux tube entanglement and carry some of the flux tubes with them. The secular spin-down of the core superfluid forces the flux tubes to move toward the stellar equator. Since the footpoints of these flux tubes are anchored in the crust, stresses grow eventually leading to failure of the crust occasionally \citep{RudermanZC1998}. As an aftermath of a quake, the global magnetic field geometry of pulsar experiences a change by the shift of inclination angle through horizontal and azimuthal motions of the broken platelet involved in the event \citep{Epstein2000}. The resulting increased curvature of the field lines close to the polar cap \citep{GeppertBMM2021} as well as seismic energy released into the magnetosphere in the form of launched Alfven waves \citep{BransgroveBL2020,YuanLBP2021} can both affect pulse profile and polarization properties. Such irreversible motion of a platelet also triggers a catastrophic avalanche of unpinned vortices which are attached to the platelet and freed after the quake. Thus, crustquakes play an important role in the intimate relation between the glitches and pulse profile changes. 

\citet{Rankin1990} conjectured that pulse width at the half maximum of the peak $W_{50}$ is related to the pulsar period $P$ and and the inclination angle $\alpha$ between the rotational and magnetic axes as:
\begin{equation}
    W_{50}=2.45^{\circ}\frac{P^{-1/2}}{\sin{\alpha}}\ .
    \label{Walpha}
\end{equation}
Across the glitch 6, the $W_{50}$ value of PSR J1048$-$5832 has decreased by $\sim0.2^{\circ}$ (see, Table \ref{profile:parameter}), which by means of Equation (\ref{Walpha}) implies an increment of amount $\Delta\alpha=0.08^{\circ}$ in the inclination angle. Any irreversible change in the inclination angle will show up in the spin-down rate with an unhealed amount of magnitude \citep{ZhaoNLT2017}
\begin{equation}
  \left(\frac{\Delta\dot\nu}{\dot\nu}\right)_{\rm per,\,shift}=\frac{\sin{2\alpha}\Delta\alpha}{1+\sin^{2}{\alpha}}, 
  \label{shitftalpha}
\end{equation}
which along with the formation of new vortex trap regions of fractional magnitude \citep{AlparCCP1996,GugercinogluA2019} 
\begin{equation}
    \left(\frac{\Delta\dot\nu}{\dot\nu}\right)_{\rm per,\,trap}=\frac{I_{\rm trap}}{I}\ ,
\end{equation}
constitutes the persistent step increases in the spin-down rate which does not relax at all. In the above expressions, $\Delta\alpha$ is the shift in the inclination angle and $I_{\rm trap}/I$ is the fractional moment of inertia of the newly formed vortex trap regions via a crustquake. 
These persistent steps play a quite prominent role in the rotational evolution of the Crab pulsar \citep{LyneJGE2015}. The slowdown of the rotation of the Vela pulsar immediately before the 2016 glitch event identified by \cite{AshtonLGP2019} was interpreted by \cite{GugercinogluA2020} as a consequence of the formation of new vortex trap regions in which physical conditions are such that vortices do not creep, so the corresponding region does not participate in the deceleration of the star. To this extent formation of high vortex density trap regions within the range $I_{\rm trap}/I\approx2\times10^{-4}-2\times10^{-2}$ by a crustquake is capable of explaining persistent step increases in the spin-down rates seen after glitches \citep{GugercinogluA2020}. 
The components of a glitch relax back towards the pre-glitch level according to \citep{GugercinogluA2020,GugercinogluGYZ2022}:
\begin{equation}
    \left(\frac{\Delta\dot\nu}{\dot\nu}\right)_{\rm g}=\frac{I_{\rm exp}}{I}\frac{\Delta\nu_{\rm g}}{|\dot\nu|\tau_{\rm exp}}e^{-t/\tau_{\rm exp}}+\frac{\Delta\dot\nu_{\rm p}}{\dot\nu}\left(1-\frac{t}{t_{\rm g}}\right)+\left(\frac{\Delta\dot\nu}{\dot\nu}\right)_{\rm per,\,shift} \ ,
\end{equation}
where $I_{\rm exp}/I$ is fractional moment of inertia of the exponentially decaying superfluid region, $\tau_{\rm exp}$ is the exponential decay time-scale, $\Delta\nu_{\rm g}$ is the total glitch magnitude, $t_{\rm}$ is the inter-glitch time, $\Delta\dot\nu_{\rm p}$ is the permanent increase in the spin-down rate associated with the decoupling of the nonlinear superfluid regions from the external braking torque via collective unpinning of vortices.
Note that, we use the terms $\tau_{\rm d}$ and $\tau_{\rm exp}$ interchangeably. In our nomenclature, $\tau_{\rm exp}$ refers to the decay timescale associated with the exponential decay of the creep regions.
The shift in the inclination angle as a result of displacement of the broken platelet during the crustquake, by Equation (\ref{shitftalpha}) implies a persistent step increase in the spin-down rate of fractional amplitude $7.86\times10^{-4}$ for PSR J1048$-$5832. The increase in the inclination angle will be reversed by the alignment torques due to the combination of the magneto-dipole radiation and the plasma supply in the magnetosphere on a substantially longer timescale \citep{PhilippovTL2014}. The broken plate size $D$ can be estimated from simple geometrical arguments as \citep{AkbalGSA2015,GugercinogluA2019}:
\begin{equation}
  \left(\frac{\Delta\dot\nu}{\dot\nu}\right)_{\rm per,\,shift}=\frac{I_{0}}{I}\cong\frac{15}{2}\sin{\alpha}\cos^{2}{\alpha}\left(\frac{D}{R_*}\right) \ ,
  \label{platesize}
\end{equation}
where $I_{0}/I$ is the change in the fractional moment of inertia of the solid part of the crust due to quake, $\alpha$ is the inclination angle, and $R_*$ is the neutron star radius. Equations (\ref{shitftalpha}) and (\ref{platesize}) lead to the estimate $D/R_*=3.73\times10^{-4}$ for the broken platelet size associated with the glitch event 6. This size is comparable to those inferred for PSR J1119$-$6127 \citep{AkbalGSA2015} and the Crab \citep{GugercinogluA2019} pulsars, suggesting a universal feature related to the breaking properties of the solid neutron star matter \citep{AlparP1985,BaikoC2018}. Note the fact that the crustquake occurred close to the neutron star surface and in the vicinity of the polar cap made it possible for its effects to be observable.

\subsection{Glitch parameters in the superfluid model}

After the initial post-glitch exponential recoveries are over, the glitch observables are related to the vortex creep model parameters through the following three simple generic equations \citep{AlparB2006}:
\begin{align}
&&    \frac{\Delta\nu_{\rm p}}{\nu}=\left(\frac{I_{\rm A}}{2I}+\frac{I_{\rm B}}{I}\right)\frac{t_{\rm g}}{2t_{\rm sd}} \ ;
&&   \frac{\Delta\dot\nu_{\rm p}}{\dot\nu}=\frac{I_{\rm A}}{I} \ ;
&&    \ddot\nu_{\rm p}=\frac{I_{\rm A}}{I}\frac{|\dot\nu|}{t_{\rm g}},
    \label{creepnuddot}
\end{align}
where $I_{\rm A}/I$ is the fractional moment of inertia of the superfluid regions that gave rise to the glitch by collective unpinning event, $I_{\rm B}/I$ is the fractional moment of inertia of the superfluid regions within which unpinned vortices moved, $t_{\rm g}=\delta\nu_{\rm s}/|\dot\nu|$ is the timescale on which change in the superfluid rotational frequency $\delta\nu_{\rm s}$ is renewed by the ongoing spin-down of the neutron star, and $t_{\rm sd}=\nu/(2|\dot\nu|)$ is the characteristic (spin-down) age of the pulsar. Equations 
(\ref{creepnuddot}) uniquely determine the fractional moments of inertia involved in glitches and naturally lead to a waiting time estimate $t_{\rm g}$ in terms of the vortex creep model. Glitches 7 and 8 have a resemblance to the 1993 double glitch observed in the Vela pulsar in which two events are separated by only 32 d \citep{BuchnerF2011}. These two events are probably part of the same unpinning avalanche with the delay caused by depletion of the farthest trap after a high enough increase in the local superfluid azimuthal velocity due to the presence of discharge initiating single free vortex line \citep{ChengAPS1988}. Therefore, we evaluate glitches 7 and 8 in PSR J1048--5832 as the relaxation behavior of the same event that occurred due to the delayed discharge between the discontiguous vortex traps. The results of the application of the corresponding set of equations to the observed values given in Tables \ref{Tab:F0F1-works} and \ref{glitch:parameter} are shown in Table \ref{model:creep}. We immediately see that the theoretical estimates $t_{\rm g}$ for the time to the next glitch are in qualitative agreement with the observed inter-glitch times $t_{\rm ig,obs}$. Also, the fractional moment of inertia of the crustal superfluid involved in the glitches is $I_{\rm cs}/I\approx3\times10^{-2}$ for both events. These include corrections regarding the crustal entrainment enhancement factor of $\langle m_{\rm n}^{*}/ m_{\rm n}\rangle\cong5.1$ due to dissipationless coupling of the dripped superfluid neutrons with the ion lattice \citep{DelsateCGFPD2016} and suggests that the maximum available superfluid angular momentum reservoir to be depleted at glitches of PSR J1048$-$5832 is around this value, thus providing an independent constraint on the equations of state (EOS).  

\begin{table}
\caption{The vortex creep model parameters found upon applying Equations 
(\ref{creepnuddot}) to the measurements for the glitch events 6 and 7. $t_{\rm ig, obs}$ is the observed inter-glitch time and $I_{\rm cs}/I$ is the fractional moment of inertia of the crustal superfluid that involved in the corresponding glitch event after corrected for the crustal entrainment effect. See the text for details.} \label{model:creep} 
    \renewcommand{\arraystretch}{1.15}
    \setlength{\tabcolsep}{5pt} 
 \begin{tabular}{cccccc}
\hline
\hline
Glitch No  & $t_{\rm ig, obs}$ & $t_{\rm g}$ & $I_{\rm A}/I$ & $I_{\rm B}/I$ & $I_{\rm cs}/I$ \\
           & (days)       &  (days)     &  $(10^{-3})$& $(10^{-3})$ & $(10^{-2})$ \\
\hline
6              &2261(19)   &2290(69)  &2.48(1) &16.6(7)  &3.13(7)\\
7                 &2456(26) &2661(405) &2.05(22) &15.6(25)  &2.60(25)   \\
\hline
\end{tabular}
\end{table}

\begin{figure}
    \centering
    \includegraphics[width=1.0\linewidth]{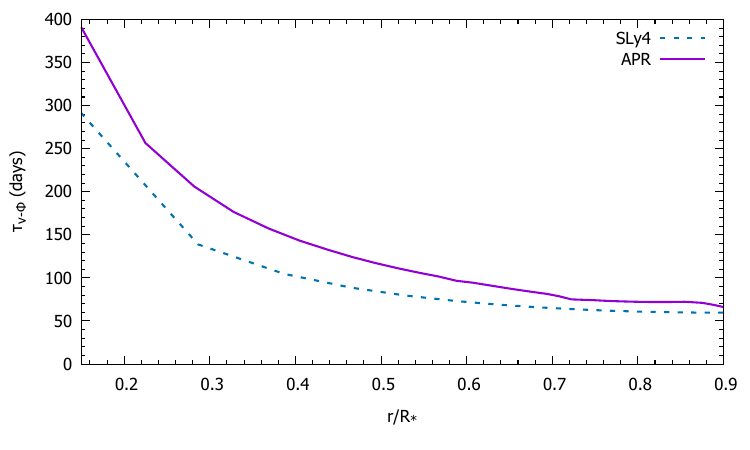}
    \vspace{-0.5cm}
    \caption{Exponential decay timescale predicted by Equation (\ref{tautor}) versus the distance from the stellar center normalized to radius for two different EOS: SLy4 (dashed blue line) and APR (solid purple line). See the text for details.}
    \label{fig:taucore}
\end{figure}

According to the vortex creep model, the response of the outer core superfluid to a glitch is exponential relaxation through vortex creep across magnetic flux tubes and the corresponding decay timescale is given by \citep{GugercinogluA2014}:
\begin{align}
 \tau_{\rm v_-\Phi}&\simeq60\left(\frac{|\dot\Omega|}{10^{-10}\mbox{
\radss}}\right)^{-1}\left(\frac{T}{10^{8}\mbox{
K}}\right)\left(\frac{r}{10^{6}\mbox{ cm}}\right)^{-1}
x_{\rm p}^{1/2}\times\nonumber\\
& \left(\frac{m_{\rm p}^*}{m_{\rm p}}\right)^{-1/2}\left(\frac{\rho}{10^{14}\mbox{\gcc}}\right)^{-1/2}\left(\frac{B_{\phi}}{10^{14}\mbox{ G}}\right)^{1/2}\mbox{days},
\label{tautor}   
\end{align} 
where $\dot\Omega$ is the slow-down rate of the star's angular velocity, $T$ is the temperature of the stellar interior, $r$ is the distance of the vortex line-flux tube junction from the center of a neutron star, $x_{\rm p}$ is the proton fraction, $\rho$ is the matter density, $ m_{\rm p}^{*}(m_{\rm p})$ is the effective (bare) mass of protons, and $B_{\phi}$ is the toroidal component of the core magnetic field. Note that for this case the glitch recovery parameter is roughly equal to the fractional moment of inertia of the core region within which vortex line-flux tube pinning interaction takes place, $Q\cong I_{\rm v-\Phi}/I$ \citep{Gugercinoglu2017}. For a representative example, we consider a $1.4M_{\odot}$ neutron star model cooling via modified Urca process \citep{YakovlevHSHP2011} and employ two different EOSs, namely SLy4 \citep{DouchinH2001} and APR (AV18 + dv + UIX*) \citep{AkmalPR1998}, for the microphysical parameters. 
The prediction of Equation (\ref{tautor}) for PSR J1048$-$5832 corresponding to the two choices of EOSs is shown in Figure \ref{fig:taucore}. 
We have identified an exponential decay timescale of 73(56) d with recovery fraction $Q=0.004(2)$ for the glitch event 7 in PSR J1048$-$5832. In previous studies on this pulsar, exponential decay terms with time scales of 160(43) d and 60(20) d and corresponding glitch recovery fractions of $Q=0.026(6)$ and $Q=0.008(3)$ were found for its two glitches occurred at MJD 49034(9) and MJD 50791(5), respectively \citep{WangMPB2000,YuMHJ2013}. Given that for the range of radii $r\cong(0.3-0.9)R_{*}$ that contain sufficient moment of inertia $I_{\rm v-\Phi}/I\cong Q$ implied by post-glitch relaxation the predicted exponential recovery timescales are $\tau_{\rm v_-\Phi, APR}=(65-206)$\, d and $\tau_{\rm v_-\Phi, SLy4}=(60-139)$\, d, we conclude from Figure \ref{fig:taucore} that the relatively stiff APR EOS accounts for all available observations better.
More detailed studies on EOS with more gamma-ray pulsars will be reported in a separate paper.

\begin{figure}
    \centering
    \includegraphics[width=1\linewidth]{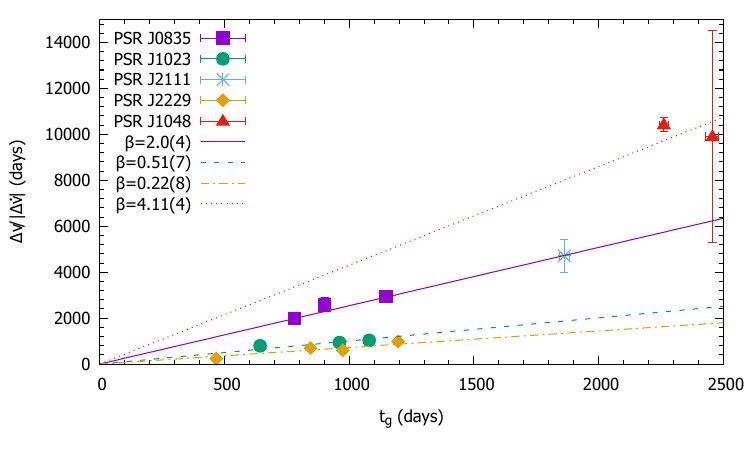}
    \vspace{-0.5cm}
    \caption{The correlation predicted by Equation (\ref{creepcor}) for PSR J1048$-$5832 (this work) and its comparison with four gamma-ray pulsars from \citet{GugercinogluGYZ2022}.}
    \label{fig:creepcor}
\end{figure}

Vortex creep model predicts a correlation between the ratio of permanent (i.e. after initial exponential recoveries are over) glitch magnitudes in frequency and spin-down rate $\Delta\nu_{\rm p}/\Delta\dot\nu_{\rm p}$ and inter-glitch time $t_{\rm g}$ as given by \citep{AlparB2006,GugercinogluGYZ2022}:
\begin{equation}
\frac{\Delta\nu_{\rm p}}{\Delta\dot\nu_{\rm p}}=\left(\frac{1}{2}+\beta\right)t_{\rm g}\ ,
\label{creepcor}
\end{equation}
from Equations \ref{creepnuddot}. 
The correlation coefficient gives a measure of $\beta=I_{\rm B}/I_{\rm A}$, which is a structural parameter associated with the radial extension of the vortex trap network throughout the neutron star crustal superfluid. The results for the glitch events 6 and 7 in PSR J1048$-$5832 obtained from Equation (\ref{creepcor}) and Table \ref{model:creep} along with those for four gamma-ray pulsars taken from \citet{GugercinogluGYZ2022} are shown in Figure \ref{fig:creepcor}. PSR J1048$-$5832, marked with a red triangle, has a slightly larger $\beta$ value compared to other Vela-like gamma-ray pulsars. This is because PSR J1048$-$5832 still produces vortex traps with crustquakes that cause persistent shifts in the spin-down rate. The young pulsar PSR J0537$-$6910 is the pulsar with the highest $\beta$ value with 13.9(6) \citep{AkbalGA2021}, and the underlying physical reason for the negative braking index measurement of $n \cong -1.2$ obtained in many studies \citep{MarshallGMWZ2004,AntonopoulouEKA2018,HoEAE2020,HoKEG2022} is the long-term unhealed increases in the spin-down rate following the glitches that occur approximately thrice a year. We therefore expect PSR J1048$-$5832 to display slightly larger $\Delta\dot\nu/\dot\nu$ in the future, because crustquakes would lead to glitches with more vortices participating in the traps.


\section{Summary}{\label{Summary}}

In this paper, we study timing observations of PSR J1048$-$5832 for the range MJD 54704 -- 59977 from Fermi-LAT and the range MJD 54302 -- 59182 from Parkes.
 By synergizing data from Fermi-LAT and Parkes, we meticulously investigate both glitch events and changes in the pulse profile of PSR J1048$-$5832.
Our analysis reveals a total of 5 glitch occurrences within the timespan of nearly 16 years, including the detection of a new small glitch at MJD 56985(9). 
The fitted glitch size is $\Delta \nu/\nu \sim  2.8(5)\times10^{-9}$, and the change in the frequency derivative is found to be $\Delta \dot{\nu}/\dot{\nu} \sim  -1.3(9)\times10^{-4}$.
This is the glitch event with the smallest glitch amplitude detected in PSR J1048$-$5832 so far. 
Furthermore, our investigation provides an update of the glitch parameters for the previously reported glitches. 
By studying the evolution of $\dot{\nu}$ with time, we find that a stable $\ddot{\nu}$ appeared after glitches 6 and 8, which enabled us to calculate interglitch braking indices as 30.5(2) and 28.3(2), respectively.

For the radio band, our approach initially involves updating the ephemeris of the raw files pertaining to PSR J1048$-$5832. This realignment served to synchronize phases within each file, facilitating subsequent analysis. 
We proceed by deriving integrated averaged pulse profiles corresponding to different periods.
Then, we find an alteration in the integrated mean profile after glitches 6, 7 and 8.
Meanwhile, after these three glitches, the five Gaussian components of the integral pulse profile all show different changes.
Although no correlation has been found between pulsar spin and radiation, 
the possibility remains that alterations in the radio pulse profile could indeed be influenced by the occurrence of glitches.
Importantly, we also do not rule out that the changes in the integral pulse profile after glitches 6, 7 and 8 may be affected by mode changes.
For gamma-ray band, no significant changes in the pulse profile were observed after the glitches.
We posit that these variations in the profile of radio pulses could potentially stem from the triggering of glitches through mechanisms involving crustquakes.

Specifically, the following physical scenario was considered for the relationship between the change in the pulse profile and the glitch. Stress on the neutron star crust exerted by magnetic flux tubes in the core superfluid/superconductor in the course of radial expansion of vortex lines exceeds the yield point of the solid crust. The direction of the stress dominates towards the stellar equator, thus the broken platelet moves accordingly to relieve strain. As the platelet displaces horizontally the magnetic field lines attached are shifted so that the magnetic inclination angle undergoes a small increase irreversibly. If the platelet is close enough to the polar cap where radio emission originates, then pulse profile variation and polarization level changes are expected due to induced local curvature in the magnetic field lines' configuration. In concurrent with these modifications to the pulsar's emission signature a spin-up glitch occurs as a result of large-scale unpinning of vortex lines triggered by the motion of the broken platelet. The post-glitch recovery takes place as a result of the evolution of perturbations to the vortex creep process introduced by the glitch event. 
We have applied this model to the observations of PSR J1048$-$5832. 
We find that $\sim0.2^{\circ}$ decrement in the core component of the pulse profile across glitch 6 implies a shift in the inclination angle by an amount $\Delta\alpha=0.08^{\circ}$, which in turn gives the estimate $D\cong4$\,m for the broken platelet size. We identify an exponential decay timescale of 73 d associated with the glitch event 7, which is consistent with the response of a 1.4 $M_{\odot}$ neutron star core cooling via modified Urca process. 
We have also employed the vortex creep model equations to the observables of the largest two glitches of PSR J1048$-$5832 in our sample and obtained the constraint $I_{\rm cs}/I\lesssim4\times10^{-2}$ for the fractional moment of inertia of the crustal superfluid involved in its glitches after correcting for the crustal entrainment effect. Theoretical estimates for the time to the next glitch by the vortex creep model match the observed timescales quite well. From the correlation analysis predicted by the vortex creep model, we conclude that PSR J1048$-$5832 is still in the process of formation of new vortex traps via crustquakes. 
Future better observations during the SKA era~\citep[e.g.,][]{2020MNRAS.493.3608J,2022JApA...43...81S} will provide answers to various questions related to glitches and their interconnection with the pulse emission.
 

\section*{Acknowledgements}
We are thankful to Profs. M. A. Alpar, X. Zhou, Dr. J. Liu and the XMU neutron star group for helpful discussions. 
The work is supported by National SKA Program of China (No.~2020SKA0120300), the Strategic Priority Research Program of the Chinese Academy of Sciences (Grant No. XDB0550300), the National Natural Science Foundation of China (Grant Nos.~12273028 and~12041304), and the Major Science and Technology Program of Xinjiang Uygur Autonomous Region (Grant No. 2022A03013).
EG acknowledges support from the National Natural Science Foundation of China (Grant No. 11988101).
{Z.Z. is supported by the Natural Science Basic Research Program of Shaanxi (Program No. 2024JC-YBQN-0036).
W.W. is supported by Zhejiang Provincial Natural Science Foundation of China under Grant No. LQ24A030002.}
P.W. acknowledges support from the National Natural Science Foundation of China under grant U2031117 and the Youth Innovation Promotion Association CAS (id. 2021055).
S.D. is supported by the Guizhou Provincial Science and Technology Foundation (No. ZK[2022]304) and the Scientific Research Project of the Guizhou Provincial Education (No. KY[2022]132).
D.L. is supported by the 2020 project of Xinjiang Uygur autonomous region of China for flexibly fetching in upscale talents.
We acknowledge the use of the public data from the Fermi-LAT data archive.
The Parkes radio telescope is part of the Australia Telescope National Facility which is funded by the Commonwealth of Australia for operation as a National Facility managed by CSIRO. This paper includes archived data obtained through the CSIRO Data Access Portal.
\section*{Data Availability}

The data underlying this article will be shared on reasonable request to the corresponding authors.

\bibliographystyle{mnras}
\bibliography{J1048-5832} 

\bsp	
\label{lastpage}
\end{document}